\documentclass[12pt, showpacs]{revtex4}

\usepackage{amsfonts,amssymb}
\usepackage{amsmath}

\usepackage{graphicx}

\newcommand{\field}[1]{\mathbb{#1}}
\newcommand{\BC}{{\field C}}

\newcommand{\BZ}{{\field Z}}

\newcommand{\CCs}{\hbox{\ixss C\kern-.4emI}}
\newcommand{\ZZs}{\hbox{\ixss Z\kern-.4emZ}}

\newcommand{\tr}{\hbox{tr}}
\newcommand{\ket}[1]{|#1\rangle}
\newcommand{\vac}{\ket 0}

\newcommand{\vev}[1]{\langle #1 \rangle}

\begin{document}
\title{Aspects of emergent geometry in the AdS/CFT context}

\author{David Berenstein, Randel  Cotta}

\address{Department of physics,  University of California at Santa Barbara}
\begin{abstract}
We study aspects of emergent geometry for the case of orbifold superconformal field theories 
in four dimensions, where the orbifolds are abelian within the AdS/CFT proposal. In particular, we show that the realization of emergent geometry starting from the ${\cal N}=4 $ SYM theory in terms of a gas of particles in the moduli space of vacua of a single D3 brane in flat space gets generalized to a gas of particles on the moduli space of the corresponding orbifold conformal field theory (a gas of D3 branes on the orbifold space). Our main purpose is to show that this can be analyzed using the same techniques as in the ${\cal N}=4$ SYM case by using the method of images, including the measure effects associated to the volume of the gauge orbit of the configurations. This measure effect gives an effective repulsion between the particles that makes them condense into a non-trivial vacuum configuration, and it is exactly these configurations that lead to the geometry of $X$ in the $AdS\times X$ dual field theory.

\end{abstract}


\maketitle

\tableofcontents

\section{Introduction and outlook}

Recently, a new program has been started to explain the origins of the AdS/CFT correspondence \cite{M}  for  four dimensional superconformal field theories (SCFT's) based on ideas of collective phenomena and a systematic expansion of the field theory around all BPS configurations of the conformal field theory compactified on an $S^3$ \cite{B}. Via the operator state correspondence, these BPS states correspond to the chiral ring of the SCFT. 

The purpose of this paper is to extend these ideas further to the study of some abelian orbifold geometries of the form $\BC^3/\BZ_n\times \BZ_m$, a step beyond work done in \cite{BCorr} ( that is a special case of orbifolds
$\BC^3/\BZ_n\times \BZ_n$ with discrete torsion). Our purpose is to show that the same techniques that work for ${\cal N}=4$ SYM are applicable uniformly to all of these examples if one uses the method if images judiciously. To explain how this will work, we need to review the 
program initiated in \cite{B} towards understanding how geometry emerges from field theory.

The important new insight of \cite{B} is that one needs to study the field theory in a semiclassical regime for all the BPS configurations, and not just small perturbations around the vacuum. This is very different in spirit from algebraic manipulations of operators, as is usually done in perturbative evaluation of anomalous dimensions of local operators.

For the case of ${\cal N}=4$ SYM theory, this BPS classical regime turns to be equivalent to studying configurations of constant commuting matrices on the $S^3$ and hence, one naturally lands in the moduli space problem of the ${\cal N}=4 $ SYM theory. For some of the deformations of ${\cal N}=4 $ SYM (in particular the so called $q$-deformation), this analysis requires to solve the F and D-terms of the 
theory, and is naturally associated to the classical moduli space problem of the conformal field theory as well (this turns out to be true for many SCFT's, although for those that do not have a free field limit the argument that needs to be made is much more sophisticated \cite{Binprogress}).

The classical BPS problem becomes the determination of BPS trajectories on the moduli space of vacua of the field theory, and these just correspond to the holomorphic fields having a very simple time dependence of the form
\begin{equation}
\phi(t) = \phi(0) \exp( it)
\end{equation}
In the zero coupling limit, this particular time dependence is  almost guaranteed because the fields $\phi$ couple
conformally to the metric of $S^3$, and one has a free harmonic oscillator of the required frequency. However, the fields $\phi$ are complex and in general they have two polarizations (one with dependence $\exp(it)$ and the other one with
dependence $\exp(-it)$), so we set the second polarization component to zero amplitude.
Thus we are choosing a circular polarization for the corresponding two dimensional (holomorphic) harmonic oscillators.

From the point of view of quantum mechanical oscillators, the dependence of $\exp(it)$ arises from  writing  the constant mode of $\phi$ on the sphere in terms of raising and lowering operators as
\begin{equation}
\phi \sim a_{\phi}^\dagger + a_{\bar \phi}
\end{equation}
so that a classical field $\phi$ as above can be understood as a coherent state 
\begin{equation}
\exp( A_{\phi} a_{\phi}^\dagger) \vac
\end{equation}
where the raising oscillators for $\bar \phi$ are not acting on the ground state. This is why these states can also be thought of as arising by linear combinations of the form
\begin{equation}
\phi^n \vac
\end{equation}
and one can relate these types of states to elements of the chiral ring, via the operator state correspondence.   These elements of the chiral ring are putatively of the form $\tr(\phi^n)(0)$, or multi-traces. The fact that multi-traces are protected operators goes back to \cite{CJR}.

The introduction of traces in the quantum oscillators results as described above is required from imposing gauge invariance.
Near the free field limit, one can understand that all the states in the chiral ring
correspond to those states as described above. These satisfy the BPS bound in the free field limit and their one-loop anomalous dimension vanishes (this problem generically leads to a spin chain model if one can study it in the planar limit \cite{MZ}). The semiclassical description above for the case of half BPS states produced a dynamics that was equivalent to an integer quantum hall droplet in the lowest Landau level \cite{Btoy}, and the geometry of these droplets in the plane
was identified exactly with the description of supergravity geometries in \cite{LLM}.

The second important issue for the techniques developed in \cite{B}, is that we can diagonalize the matrices $\phi$ simultaneously for all of these BPS configurations by gauge transformations, because they commute for the relevant configurations. This is analogous to the case of half-BPS states, where the relevant dynamics consists of a single Hermitian matrix quantum mechanics problem, and it's canonically conjugate momentum. The diagonal (eigenvalue) degrees of freedom are what gives
rise to the coordinates of the free fermions in the half BPS case.

Based on this observation, one is really describing the moduli space of vacua problem of the theory, and not just the set of all solutions of the $D$-term and $F$-term constraints without imposing the gauge equivalence of configurations. This diagonalization is the same type of problem that in M(atrix) theory leads to consider $N$ D0 branes (particles) at particular locations (determined by the eigenvalues) in ten dimensions \cite{BFSS}. Thus the moduli space problem that we are interested in is given roughly by $N$ particles on the moduli space of vacua of a single brane in the bulk. 
This also works in the case of $q$-deformations of ${\cal N}=4 $ SYM theory. We will name these eigenvalue-like variables that describe the positions of these D-branes as {\em diagonal variables}, although we will in see in the examples in this paper that they are not strictly given by diagonal matrices, but rather block-diagonal collections of matrices.

If we reduce the system classically to these configurations where the $F$ and $D$ terms vanish, we get an effective gauge  dynamics on these sets of these configurations, and we will assume that this effective dynamics describes exactly the dynamics of the BPS configurations.
If we write the set of configurations in terms of a diagonal system of variables, we get
exactly $N$ free particles in the moduli space of vacua of a a single D-brane.

In quantum systems, making a change of variables gives rise to changes in the measure, and they affect the Hamiltonian of the theory when  compared with the classical result. This 
can be quantified by the volume of the gauge orbit of a particular configuration and gives an effective repulsion between the eigenvalues.

The effective dynamics for this problem was solved in particular cases in \cite{B, BCV,BCorr}. The end result is that the effective repulsion of the eigenvalues gives rise to a non-trivial distribution of particles (eigenvalues) in the ground state and the excited states. In the large $N$ limit (which is a thermodynamic limit for these particles), one can take a hydrodynamic approximation to the problem and one can change variables further to the density distribution of these particles in the moduli space of vacua.
It is these distributions of density that can be correlated directly with geometry. Moreover, in the large $N$ limit the distribution density can be calculated using a saddle point approximation for all the states one is interested in \cite{B}.

For ${\cal N}=4 $ SYM, the ground state distribution is a five-sphere whose radius is of order ${\sqrt N}$ (the exact radius depends on the choice of normalizations of fields. This was calculated precisely in \cite{BCV}). Writing effective coherent states using the known map between supergravity states and fields \cite{M,Wads,GKP}, one finds that this sphere gets deformed exactly according to the spherical harmonic of the
corresponding supergravity mode. In principle one would also want to check non-linear effects on this distribution to calculate OPE coefficients at large t' Hooft coupling to make sure that one understands the relative normalization of these deformations, but this has not been done yet.

For the time being, it seems that this distribution of eigenvalues coincides exactly with the geometry of the sphere of the supergravity limit of the dual theory. This same reasoning works for the case studied in \cite{BCorr}. Finally, one realizes that after the size of the distribution 
of particles is taken into account, off-diagonal modes become very massive, with a typical mass of order 
$\sqrt{g^2N}$ and hence one can also think of the BPS moduli space approximation as 
a Born-Oppenheimer expansion where heavy modes have been integrated out. Indeed, one can use single excitations of these off-diagonal modes to calculate exactly the spectrum of BMN energies \cite{BMN} at
large values of the 't Hooft coupling $\lambda= g_{YM}^2 N$ directly from the field theory \cite{BCV}. The result of this calculation coincides with the 
two loop explicit calculation of \cite{GMR}, and with the all loop calculation \cite{SZ}.  The calculation in \cite{BCV} and \cite{SZ} rely on completely different arguments. This result also matches the required energies 
for the all loop Bethe ansatz conjecture of \cite{AFS,BDS}, and it can be tied to an enhanced algebraic structure of the planar anomalous dimension problem \cite{Beis}, where the set of supersymmetry charges on the string seems to get a central extension. This extended algebra reproduces this same result by studying short multiplets of the centrally extended symmetry algebra.

More recently, it has been shown in \cite{HM} that the result of these energies beyond the BMN limit provided in \cite{SZ,BCV} can also be reproduced (these off-diagonal modes become the so called giant magnons in the Nambu-Goto string propagating on $AdS_5\times S^5$). The calculations in \cite{HM}  also match the  basic features of the geometry of the string derived from field theory considerations, that appeared from considering off-diagonal modes as {\em string bits} in \cite{B, BCV}. In the description above, off-diagonal modes to linearized order join two specific eigenvalues (these become points in the density distribution of particles in the moduli space). To build gauge invariant states, one has to make closed paths in the eigenvalue density distribution, by connecting different eigenvalues with (straight) lines. One gets in this way a polygonal description of the string, where the edges of the polygon are these off-diagonal modes. It is natural to think of these as string bits. so one has a basic geometric picture in the field theory description of where the strings are in the geometry. These states get dressed by motion of the eigenvalues themselves, so one gets something like a gravitational dressing of the string bits to make a full string. 

The purpose of this paper is to extend this type of geometrical picture to other setups. In particular we are interested in the description of abelian orbifolds of $AdS_5\times S^5$, associated to D-branes on the orbifold $\BC^3/\Gamma$, with $\Gamma$ an abelian group.
The main issue we want to address is if the method of images can be used throughout to make calculations. In this sense, all the results from $N=4$ SYM could be carried over with little modification, and this would basically encode the fact that the dual geometry is exactly $AdS_5\times S^5/\Gamma$ thought of as an orbifold.

The great advantage of studying these models, is that their field theory duals are very well known and have been studied a lot, beginning with \cite{KS}.  

Since it is already well known that for orbifolds one can use the method of images to compute the off-diagonal mode spectrum of strings stretching between branes, and that in general planar diagram amplitudes are inherited from ${\cal N}=4 $ SYM, the main question that is still left to be answered  within the techniques described above, is if we can also use the method of images to get the effective repulsion between eigenvalues at strong coupling due to the quantum measure effects. With this repulsion we can just go to the covering space and we recover the 
distribution of particles associated to ${\cal N}=4 $ SYM,  so the geometry that the strings bits would see would exactly coincide with the idea of an orbifold.

 The final answer is that the method of images can be used and our paper provides a very detailed explanation of how this happens. Our techniques in principle also have applications to 
other problems related to matrix models. 
This gives us confidence
that the method of images is very robust for all cases of interest, although we do not have a
general proof that is applicable for all supersymmetric examples yet. Moreover, understanding how this method of images  works in practice enables us to give a coherent setup for calculations associated to field theories that are not orbifolds. It should also be clear that for the most part our techniques also generalize to the case of orientifolds, so long as one is careful with the handling of the gauge group analysis.

For interesting generalizations to non-orboifold setups, one could consider
deforming an orbifold theory by changing the ratios between the gauge couplings. Some such deformations are marginal and lead to families of conformal field theories that are not strict orbifolds, although they are very closely related to orbifolds.  It will be clear from our calculations that this deformation should affect the volume of the gauge orbit and this might have important consequences for understanding the string
theory duals to such deformations further.
 Working the details for non-orbifold examples is beyond the scope of the present
paper. However, the general setup we provide can be easily generalized to those other 
examples although most calculations become a lot more involved in general. These are certainly interesting avenues to explore for the future. Another example that is very interesting to explore is a particular case of the general Leigh-Strassler deformations of ${\cal N}=4$ SYM 
\cite{LS}. These deformations lead to interesting geometries \cite{D, BL,LM}, a special case of which has already been studied with the techniques we are using here \cite{BCorr}. In \cite{Brev} another interesting family was 
found that  leads to an interesting geometry on the  moduli space of vacua which is not of the  orbifold type. These particular theories have also been studied recently using various perturbative techniques \cite{BM,MPPSZ}. These examples discussed here are important because they can be realized in terms of a CFT with a free field limit. In this case understanding what happens when  we dial the 't Hooft coupling will give us a clue as to how $\alpha'$ corrections work in the AdS/CFT and we need ways to interpolate between weak and strong coupling systematically.

\section{  The field theory}{\label{sec:quiver}}

The field theory associated to D-branes near the tip of a supersymmetric orbifold singularity $\BC^3/\Gamma$ can be constructed easily using quiver diagrams and the techniques developed in Douglas and Moore \cite{DM}. These theories give rise to four dimensional conformal field theories \cite{KS} and are an interesting area to study examples of the AdS/CFT correspondence.
A complete list of the allowed sets of $\Gamma$ that are compatible with supersymmetry is available in \cite{HH}. We are interested in the particular case where $\Gamma$ is abelian, as this is a simpler example to analyze. The most general form for the a discrete abelian group of $SU(3)$ is as a product $\Gamma=\BZ_n\times\BZ_m$. This happens because $SU(3)$ is of rank two. 

To study the geometry more explicitly, we choose a set of coordinates $x,y,z$ in $\BC^3$ such that the action of $\Gamma$ is just multiplication of $x,y,z$  by phases. We can always do this because an abelian subgroup of $SU(3)$ can be conjugated to lie in the Cartan subgroup, and under the Cartan subgroup one can diagonalize the action to get different charges under $U(1)^2 \subset SU(3)$.

The quiver diagram takes the form of a grid of $n\times m$ nodes. One in general gets
a brane box model, like \cite{HZ,HU}.  Nodes in such a diagram are associated with all the inequivalent irreducible representations of $\Gamma$ which can be labelled by coordinates $(\alpha,\beta)$ with $\alpha+n\equiv\alpha$ and $\beta+m\equiv\beta$. The translation moves along the quiver correspond to multiplying a given representation by a generating character of $\BZ_n$ and $\BZ_m$ respectively.

Arrows between nodes are associated with chiral superfields which map between inequivalent representations by multiplication by a character of the orbifold group $\Gamma$. These characters are exactly the phases under which the coordinates $x,y,z$ transform. These fields transform naturally as $x,y,z$ and this fixes the set of nodes that a field is allowed to connect.

In the quiver, this gives us $n m$ fields originating from a single field $X,Y$ or $Z$ from ${\cal N}=4$ SYM.
If we label nodes by $(\alpha,\beta)$, we can take the fields joining two nodes and use the nodes as labels, e.g.
\begin{equation}
X_{(\alpha,\beta),(\alpha+1,\beta+s)}
\end{equation}
We will formally sum these into a field $X$
\begin{equation}
X= \oplus_{\alpha,\beta} X_{(\alpha,\beta),(\alpha+1,\beta+s)}
\end{equation}
Indeed, it was found in \cite{Brev} that such formal sums were a convenient algebraic manipulation that lets us write the F-term relations in terms of the $X,Y,Z$ defined above in a simple way.

To do this, one also associates to each node a projector $P_{\alpha,\beta}$, so that we can recover the individual fields as
$X_{(\alpha,\beta),(\alpha+1,\beta+s)}$ by acting with $P_{\alpha,\beta}$. This is,
\begin{equation}
X_{(\alpha,\beta),(\alpha+1,\beta+s)} = P_{\alpha\beta} X
\end{equation}
and a similar relation from multiplication on the right by $P_{\alpha,\beta}$.

The complex conjugate fields that are antichiral have ``arrows" in the opposite direction.
These will be important to check the D-term constraints.

Also, associated to each vertex, we will have a gauge group $U(N_{\alpha\beta})$. The arrows coming into the node are in the antifundamental, while those leaving the nodes are in the fundamental. 
The important thing for us, is that to make gauge invariants we need to contract fundamentals with antifundamentals. This becomes just matrix multiplication of arrows. For example $X_{(\alpha,\beta),(\alpha+1,\beta+s)}$ can be concatenated on the right with any field transforming as
\begin{equation}
T_{(\alpha+1,\beta+s),(\gamma,\delta)}
\end{equation}
We do the same with the gauge fields and define
\begin{equation}
V =\oplus V_{(\alpha,\beta), (\alpha,\beta)}
\end{equation}
with the vector superfields. 

The description given here makes it natural to work with the path algebra of the quiver diagram
(this has been explained in  \cite{Brev} ). For our purposes we are interested in holomorphic invariants to describe the Moduli space of vacua of the theory, so we only need $X,Y, Z$ and the $P$ to do most of our work. On the moduli space, the F-term and D-term equations are satisfied. One usually finds first solutions of the F-term constraints, and then one can impose the D-term constraints later on. These F-term equations are relations in the quiver path algebra.

The field theory on flat space can be written identically to the ${\cal N}=4 $ SYM theory
\begin{equation}
S = \frac 1{g_{YM}^2}\tr \left[\int d^4 \theta \bar X \exp(V)X \exp(-V)+ \left(\int d^2\theta X[Y,Z]+ W^2\right) + c.c. \right]+ \hbox{$\theta$-terms}
\end{equation}
if we remember that $X,Y,Z$ are not completely generic, but that they can be written exactly as
\begin{equation}
X= \sum_{\alpha,\beta} P_{\alpha,\beta} X P_{\alpha+1,\beta+s}
\end{equation}
and similarly for $Y,Z,V$. Our notation is that $X$ shifts the lattice by $(1,s)$, $Y$ will shift it by 
$(\ell,r)$ and $Z$ has to shift it by $(-1-\ell, -s-r)$. Notice that all of the gauge coupling constants 
for the individual gauge groups are the same. This is required for the closed string theory
CFT on the space $\BC^3/\Gamma$ to be the orbifold field theory, and not some other CFT that can be reached by turning on twisted sector marginal deformations of the orbifold CFT.

These couplings being equal also gives an enhanced discrete symmetry of permutations of the nodes of the quiver by lattice translations. This symmetry requires all coupling constants to be the same and it is essentially the quantum symmetry of the orbifold conformal field theory.

All the superpotential relations can be cast as 
\begin{eqnarray}
XY-YX&=&0 \\
YZ-ZY&=&0 \\
ZX-XZ&=&0
\end{eqnarray}

Since the orbifold is obtained by an abelian quotient, all the irreducible representations of $\Gamma$ are one dimensional. A brane in the bulk corresponds to the regular representation of the group (this is even true in the case of discrete torsion, with appropriate definitions. See for example \cite{BL}).

To solve these equations for a single brane in the bulk, it is convenient to realize the algebra as a set of $nm\times nm$ matrices, where we use the $(\alpha,\beta)$ pairs to denote the rows of the matrices. 
We also denote the columns with double indices. This reordering makes it easy to introduce 
elementary matrices
\begin{equation}
E_{(\alpha,\beta), (\gamma,\delta)}
\end{equation}
with multiplication rule
\begin{equation}
E_{(\alpha,\beta), (\gamma,\delta)}\cdot E_{(\alpha',\beta'), (\gamma',\delta')}
= \delta_{(\gamma, \delta),(\alpha',\beta')}E_{(\alpha,\beta), (\gamma',\delta')}\label{eq:Ecomm}
\end{equation}
A general solution to the above equations (up to gauge equivalence) is given by
\begin{eqnarray}
X&=& x\sum_{\alpha,\beta} E_{(\alpha,\beta)  (\alpha+1,\beta+s)}\\
Y&=& y \sum_{\alpha,\beta} E_{(\alpha,\beta)  (\alpha+\ell,\beta+r)}\\
Z&=& z \sum_{\alpha,\beta} E_{(\alpha,\beta)  (\alpha-1-\ell,\beta-s-r)}
\end{eqnarray}
for $x,y,z$ arbitrary complex numbers
and $P_{\alpha,\beta}= E_{(\alpha,\beta),(\alpha,\beta)}$.
A general gauge transformation will be a general element of $U(1)^{mn}$, and it is encoded as
\begin{equation}
G = \sum \exp( i \theta_{\alpha,\beta})P_{\alpha,\beta}
\end{equation}
In this notation $G$ acts on the fields by adjoint action (this is how the bifundamental constraint gets implemented), this is $X\to G X G^{-1}$. From here, $x,y,z$ are not gauge invariant. Instead, only combinations of $X,Y,Z$ that give rise to closed paths in the quiver give rise to gauge invariant variables, these are certain monomial combiations of $x,y,z$.

The interpretation of $x,y,z$ is as the coordinates of one of the preimages of the D-brane in the covering space $\BC^3$. Since they are just numbers, it is obvious that $xy-yx=0$ etc.
Thus, these represent a point in the moduli space of vacua of the ${\cal N}=4 $SYM theory. 
One can find gauge transformations that multiply $x,y,z$ by global phases. These gauge transformations permute between the different preimages of the brane in the orbifold quotient.
The monomials that are described above as being gauge invariant are just the holomorphic invariants of the orbifold action itself.

Notice that with the above solution, we have also automatically solved the D-term constraints
as well
\begin{equation}
[X,\bar X]+ [Y,\bar Y] +[Z,\bar Z]=0
\end{equation}
Even better, all commutators between the fields $X,Y,Z$ and their conjugates vanish.

From this, it is clear that in this formulation $X,Y,Z$ (and their adjoints) are normal commuting matrices, and they can be diagonalized simultaneoulsy (but not by an allowed gauge transformation). One can  show easily that if we look at these collections of eigenvalues, one gets exactly all the preimages on the covering space $\BC^3$, of a point in the orbifold
$\BC^3/\Gamma$.

This is how the method of images makes it's first appearance directly in the standard orbifold construction. For orbifolds with discrete torsion, this does not work exactly in the same way, as the field configurations that give rise to the moduli space of vacua do not commute anymore \cite{D,BL,DF, BJL}. 

To get many D-branes, one just copies this solution in block diagonal form. In the work \cite{BL2,Brev}, this is done by realizing that the moduli space problem is a representation of the
quiver path algebra with $F$-term relations (we also check the D-terms afterwards). Within algebras, direct sums of representations 
are new representations, and the direct sum decomposition is the decomposition into block diagonal form alluded to above.

For us, it is interesting to write these solutions with the help of the following matrices
\begin{eqnarray}
E_X&=&\sum_{\alpha,\beta} E_{(\alpha,\beta)  (\alpha+1,\beta+s)}\\
E_Y&=&\sum_{\alpha,\beta} E_{(\alpha,\beta)  (\alpha+l,\beta+r)}\\
E_Z&=&\sum_{\alpha,\beta} E_{(\alpha,\beta)  (\alpha-1-l,\beta-s-r)}
\end{eqnarray}
so that the solution described above becomes
$X= x E_X$, $Y= y E_Y$, $Z= z E_Z$.

If we assume that the rank of the gauge group is a product $N=kn$ then we can construct an $N$-dimensional representation of the same algebra:

\begin{eqnarray}
\widehat{X}&=&\Lambda^1\otimes\sum_{\alpha,\beta} E_{(\alpha,\beta)  (\alpha+1,\beta+s)}= \Lambda^1\otimes E_X \\
\widehat{Y}&=&\Lambda^2\otimes\sum_{\alpha,\beta} E_{(\alpha,\beta)  (\alpha+l,\beta+r)}=\Lambda^2\otimes E_Y \\
\widehat{Z}&=&\Lambda^3\otimes\sum_{\alpha,\beta} E_{(\alpha,\beta)  (\alpha-1-l,\beta-s-r)}= \Lambda^3\otimes E_Z
\end{eqnarray}
where $\Lambda^1$, $\Lambda^2$ and $\Lambda^3$ are commuting $k\times k$ matrices (this is just imposing the commutation relations for $X,Y,Z$).  Since they are mutually commuting the $\Lambda$'s can be simultaneously diagonalized with a unitary transformation. 
This transformation is part of the gauge symmetry of the system, so it amounts to a choice of gauge. The $U(N)$ transformation that is required is embedded diagonally on all nodes of the quiver.

Working with diagonal $\Lambda$'s reduces the degrees of freedom of the system to that of their eigenvalues, and gives us the block diagonal representation that we want. Basically, we have replaced $x, y,z$ by diagonal matrices.

Finally, it is easy to compute the masses of massive fields associated to strings stretching between two D-branes in the bulk. The calculation 
gives the same result as in ${\cal N}=4 $ SYM, where we get massive particles from all straight line segments between the preimage of a single brane, and all the images of the second brane
(this has the same number of polarizations as massless representation of ${\cal N}=4 $ SYM for each such segments).
It does not matter which preimage we choose for the first brane, so long as we remember all the
other images of the second brane. There is no extra multiplicity from shuffling amongst the 
preimages of the first brane (this ends up being the way we get the orbifold projection on open string states \cite{DM}).

We would also like to mention that the spectrum of anomalous dimensions of these theories has been studied systematically at weak 't Hooft coupling in \cite{WW,SS,BR}. Those works should be considered as complimentary to the present approach.

\section{  Emergent geometry and the method of images}{\label{sec:images}}

Now that we have understood the basic structure of the field theories of interest, we want to carry out the program of \cite{B,BCorr} to determine the geometrical structure of the gravitational dual. The main idea is that for an orbifold, one should be able to use the method of images to determine or impose that the geometry is indeed that of an orbifold.

The first step, is to study the free field limit of the field theory, compactified on $S^3$.
Formally, we can do this by writing all fields in a spherical harmonic decomposition of the sphere.

Because of the operator state correspondence, these modes are in one to on correspondence
with local operator insertions at the origin. For example, the operator $\partial^{[n]} \phi (0)$
where $[n]$ is a multi-index, becomes identified with the $[n]$-th spherical harmonic of the field on $S^3$ (for no derivatives, this is the s-wave of the field).

Once we have these free fields, we want to understand what the BPS states will look like. 
This is just studying the chiral ring for our purposes. The elements of the chiral ring are
roughly of the form
\begin{equation}
\tr( X^{n_1} Y^{n_2} Z^{n_3})
\end{equation}
and products of these.
Because we have more gauge groups, we can also project onto particular gauge groups in the quiver, but this does not change the description signifcantly.
These elements don't depend on derivatives of the field, so the dynamics of BPS states reduces to constant modes on the sphere.

Moreover, the fields appearing in the operator are identified with the raising operators of the corresponding quantum field on the sphere. Thus a state as described above, becomes 
in the classical limit a classical trajectory on the space of constant matrices with circular polarization. This gives us a specific time dependence for $X,Y,Z$ of the form
\begin{equation}
X(t)= X(0) \exp( it), Y=Y(0) \exp(it), Z(t) = Z(0) \exp (it) \label{eq:BPSclass}
\end{equation}
where in the above, $X,Y,Z$ are constant matrices on the sphere.

The second step is to reduce the dynamics to these classical configurations of constant matrices and to analyze the classical BPS equations with more care. The classical dynamics is 
a (multi) matrix model quantum mechanics, which is just the dimensional reduction of the field theory on a sphere.

The BPS bound of the supersymmetry algebra is that the Hamiltonian is equal to the R-charge. Because we have now a finite number of degrees of freedom, we can turn on the interactions perturbatively and analyze their effects classically.

For a solution of the type described above, in the free field limit we saturate the BPS bound.
Turning on the interactions slowly, the $F$ and $D$ terms will contribute to the Hamiltonian
by adding a positive square to it, while they will not contribute to the R-charge. The $R$ charge does not get corrected, because one can tie the results to ${\cal N}=4 $SYM where the $R$ charge is part of a non-abelian symmetry, and it's eigenvalues have a strict quantization. Classically this property means that the R-charge of a configuration does not depend on the potential.

Thus, only classical solution of the type given in equation ( \ref{eq:BPSclass}) are allowed, if $X,Y,Z$ commute with each other and the D-terms of the theory vanish.
Incidentally, setting F and D-terms to zero is exactly the moduli space problem of the corresponding field theory, which we have already described in the previous section in detail.

Thus, the full BPS dynamics is reduced to configurations of the type

 \begin{eqnarray}
X&=&= \Lambda^1(t)\otimes E_X \\
Y&=&\Lambda^2(t)\otimes E_Y \\
Z&=&\Lambda^3(t)\otimes E_Z
\end{eqnarray}
where the $\Lambda^i$ are commuting matrices.

This dynamics is a special set of configurations of constant matrix dynamics, whose classical Lagrangian is just the field theory on the sphere with constant modes (the dimensional reduction of the field theory on the sphere). This dimensional reduction is a consistent classical truncation of the dynamics of the original problem, which is characterized by being invariant under rotations on the three sphere.

The idea now is to calculate the effective quantum Hamiltonian for the BPS dynamics of the system, as a usual Schrodinger operator on the variables $\Lambda^i$. 
Since the original theory on the $S^3$ is gauged, the dimensional reduction will have the same property, and the effective dynamics is a gauged matrix quantum mechanics, where the gauge
invariance is by constant unitary transformations of the fields on $S^3$ (the gauge invariance compatible with having constant vevs for the fields $X,Y,Z$). We want to choose a gauge where the $\Lambda^i$ are diagonal, so that we are describing trajectories on the moduli space of vacua, this is, we want to focus BPS states.

In the ${\cal N}=4 $ SYM theory, this was done in \cite{B}, with the result that the effective quantum Hamiltonian was a set of $N$ harmonic oscillators in six dimensions,  with a modified Laplacian
\begin{equation}
H_{eff} = \sum_i \frac 1{2\mu^2}\nabla_i \cdot\mu^2 \nabla_i +\frac 12 \vec x_i^2
\end{equation}

In the expression above, the moduli space of vacua was characterized by the position of $N$
particles on six dimensional flat space. These are given by the label $i$. The expression $\nabla_i$ denotes the vector gradient operation associated to the coordinates of each 
particle in six dimension, and the dot product indicates that we are taking an effective Laplacian. 

If $\mu^2$ where not present, the result above is just $N$ particles in a six dimensional harmonic oscillator. This would be the classical result from using a diagonal ansatz for the problem from the start. The reason the Laplacian is not exactly the classical result, is that on going to eigenvalues, we have made a gauge choice, and we have gone to a coordinate system of the matrices that is more akin to spherical coordinates than to flat coordinates. Thus, the Laplacian has a non-trivial measure attached to it.

After all, all the solutions we are interested in to understand the full quantum mechanics of the moduli space associated to ${\cal N}=4 $SYM are of the form
\begin{equation}
X^i = U X^i_{diag} U^{-1}\label{eq:gaugetr}
\end{equation} 
and not just the diagonal ones.

The matrix $U$ is characterized by some angular coordinates, so it is interesting to write the kinetic term of $X^i$ in terms of it's eigenvalue times derivatives, and the time derivatives of the angles appearing in $U$. (This decomposition is illustrated very nicely in section 3 of the review \cite{Kleb})

This kinetic terms is obtained from
\begin{equation}
\frac 12 \tr(\dot X)^2 
\end{equation} 
The kinetic term that one obtains using this method, near the identity $\theta=0$, is of the form
\begin{equation}
\frac 12 (\dot X_{diag})^2+ f(X_{diag})_{\alpha\beta} \dot \theta^\alpha \dot \theta^\beta
\end{equation}
This is a special form of the metric. In general one would expect that the angles $\theta$ and the directions $X$ could mix. These cross terms make calculations more difficult.
Part of our goal is to show that a similar result holds in the orbifold case, so that there is no such mixing.

The Laplacian associated to such a kinetic term is
\begin{equation}
\frac 12\mu^{-2} \nabla_i \cdot \mu^2 \nabla_i + f^{\alpha\beta}\partial_\alpha\partial_\beta
\end{equation}
where $\mu^2 =\det( f_{\alpha\beta})$.
If $X$ is a unique Hermitian matrix, then $\mu^2$ is the square of the Van-der-Monde 
determinant \cite{BIPZ}. For the case of ${\cal N}=4 $ SYM, we have six commuting matrices, and $\mu^2$ has a similar form to a Van-der-Monde determinant\cite{B}.

For such a system, a wave function $\Psi$ will be gauge invariant if $\partial_\alpha \Psi=0$
for all $\alpha$. Thus the effective Hamiltonian reduces to computing the measure term
$\det(f_{\alpha\beta})$, as the second term will not matter. This determinant is a Faddeev-Popov determinant for the gauge where $X$ is diagonal. Additionally, permutations of the eigenvalues are part of the gauge transformations, so the wave function in terms of the $\vec x_i$ has to be symmetric under exchange of particles.

Choosing $U = 1+i\delta\theta_\alpha T^{\alpha}$ in equation (\ref{eq:gaugetr}) with $T^a$ suitably chosen, leads immediately to 
\begin{equation}
\mu^2 = \prod_{i <j} |\vec x_i-\vec x_j|^2
\end{equation}
This just arises from
\begin{equation}
g (d\theta)^2 \sim \tr (\delta_\theta X^i\delta_\theta X^i) \sim \tr( [X^i,d\theta]^2)
\end{equation}

The expression $\mu^2$ is induced from the volume of the orbit of a configuration in the space of fields under gauge transformations.

This same type of calculation was done in a more complicated case in \cite{BCorr}, where instead of studying the ${\cal N}=4$ SYM, a particular $\Gamma=\BZ_n\times \BZ_n$ orbifold with discrete torsion was needed. The main modification was that $\mu^2$ was different.   The expression $\mu^2$ could be calculated by the method of images. This is, the measure terms also included a product over the images of particle $j$ under the  $\BZ_n\times \BZ_n$ symmetry, and there was also a contribution from the self-images of a particle under the discrete group of identifications in the covering space $\BC^3$. Thus the formula above was modified to
\begin{equation}
\mu^2 =\left[ \prod_i \prod_{\gamma\in \Gamma, \gamma\neq 1}|\vec x_i -\vec x^\gamma_i|^2\right]\left(\prod_{\gamma\in \Gamma}\prod_{i <j} |\vec x_i-\vec x^\gamma_j|^2\right)\label{eq:images}
\end{equation}
where for each particle in the orbifold, we choose a single
preimage on the covering space $\BC^3$, $\vec x_i$, and we calculate with this preimage 
in the above formula. The choice we make does not matter.
The other necessary modification, is that there is some residual gauge symmetry of the 
$x_i$ in the choice of the preimage. The quantum wave function of the $x_i$ should be invariant under
these permutations. 

Obviously, we want to show that the result embodied in equation (\ref{eq:images}) is valid for all orbifolds of interest in this paper.

The final step in determining the geometry is to write a wave function for the eigenvalues that 
solves the associated Schrodinger problem, with $\mu^2$ included.
The gaussian wave function is such a wave function, and it defines the ground state. The width of the gaussian is determined by plugging the ansatz into the classical Hamiltonian. This usually gives factors of $|\Gamma|$, the order of the group, in various places.

Given such a wave function, the geometry in the AdS dual  is determined by finding what the distribution of eigenvalues looks like.
The distribution of eigenvalues depends on calculating integrals with the measure $\mu^2$ from the wave function. This is
\begin{equation}
\vev{f(X)} = \int f(X) \mu^2 \Psi^*\Psi\times \langle \Psi |  \Psi\rangle^{-1}
\end{equation}

Thus, it is more convenient to work with modified wave functions given by
\begin{equation}
\hat\psi = \mu \Psi
\end{equation}
Also, for the ground state wave function 
\begin{equation}
|\hat\psi|^2 \sim \exp( - \sum_i A \vec x_i^2+ \sum_{i\neq j, \gamma\in \Gamma}\log(|\vec x_i-\vec x^\gamma_j|) +\sum_{i, \gamma\in \Gamma, \gamma\neq 1} \log(|\vec x_i -\vec x_i^\gamma|^2)
\end{equation}
Now we can realize that this is the same as the Gaussian wave function one would have 
written if we used the particles and their images as if they were independent objects, except that the configuration is required to have some polyhedral symmetry, and one uses a measure
factor that is calculated using the method of images.

This system can then be analyzed exactly as in \cite{B,BCV,BCorr} and one sees that the full geometry is just the orbifold of $S^5$. This is because different images of particles are gauge-equivalent to each other, as was explained in the previous section. 

All we need to do to have this interpretation, is to be able to reproduce the measure formula
(\ref{eq:images}) for the cases of interest. This requires computing the corresponding Faddeev-Popov determinant (the volume of the gauge orbit) very carefully.

\section{Calculation of the volume of the gauge orbit}

In the previous section, we have argued what is the expected result for the volume of the gauge orbit that we want to analyze. What we need to do now, to compute the corresponding volume of the gauge orbit, is look at the form of the 
kinetic term of the gauge theory.

This is just like in the ${\cal N}=4 $ SYM case, given by
\begin{equation}
\frac 12 \tr {\dot X}\dot{ \bar X}^2+ (X\leftrightarrow Y) + (X\leftrightarrow Z)
\end{equation}

First, we assume that the rank of the gauge group is a product $N=kn$ then we can construct an $N$-dimensional representation of the same quiver algebra:

\begin{eqnarray}
X&=& \Lambda^1(t)\otimes E_X \\
Y&=&\Lambda^2(t)\otimes E_Y \\
Z&=&\Lambda^3(t)\otimes E_Z
\end{eqnarray}
where $\Lambda^1$, $\Lambda^2$ and $\Lambda^3$ are commuting $k\times$$k$ matrices.  Since they are mutually commuting the $\Lambda$'s can be simultaneously diagonalized with a unitary transformation. Working with diagonal $\Lambda$'s reduces the degrees of freedom of the system to that of their eigenvalues but requires that we take careful account of the non-trivial measure factors that come from this gauge choice. 

The measure comes from the volume of the gauge orbits in question.  To compute these we start with the choice of diagonal $\Lambda$'s and perform an infinitesimal transformation with the generators of the gauge group.  This reduces to computing commutators of the $\widehat{X}$, $\widehat{Y}$ and $\widehat{Z}$ matrices with the gauge group generators.  Since generators act by similarity transformation on each the irreducible representations of $\Gamma$ (they are associated to the nodes of each quiver), they can be written as: $\Theta=E_{(\alpha,\beta) (\alpha,\beta)}$ in the tensor product described previously. For purposes of this calculation it is a little nicer to use the discrete fourier transformed versions of these generators:

\begin{displaymath}
\widetilde{\Theta}_{kj}=\sum_{\alpha,\beta}E_{(\alpha,\beta) (\alpha,\beta)}q^{\alpha k}w^{\beta j}
\end{displaymath}
here $q$ and $w$ are primitive $n^{th}$ and $m^{th}$ roots of unity. 

The reason why these Fourier transforms behave nicer than plugging the gauge groups one by one,  is that the permutations of the nodes that are induced by tensoring with the generators of the character group of $\Gamma$ correspond to the $\hat \Gamma= \hat\BZ_n\times\hat\BZ_m$ quantum symmetry of the orbifold. The discrete Fourier transform is thus picking the infinitesimal gauge generators with fixed $\hat \Gamma$ quantum numbers (more information can be found in \cite{BL}). If we were not aware of this fact, this permutation symmetry is a symmetry of the Lagrangian and of the solution, and diagonalizing symmetries that commute with the Hamiltonian always simplifies the form of quadratic terms in the action or Hamiltonian.

In general, the quantum symmetry is not well defined on gauge variant quantities, because we can sometimes compensate by a gauge transformation and get  different quantum numbers. Here, the advantage is that the ansatz we wrote is invariant under the permutations of the nodes (they are chosen to act only on the $\alpha,\beta$ indices), so in principle all small perturbations can be chosen to diagonalize these quantum numbers, so long as there are no residual gauge transformations of the configuration. For a single brane, there are discrete symmetries that are compatible with the ansatz, and they change $x$, $y$, $z$ by roots of unity. These can be embedded in the gauge group. This means that there is no uniform way to decide the twisting of the $X,Y,Z$
fields between different branes. Once we choose a gauge without residual symmetries this is not a problem anymore. Here, we just keep the values of $x,y,z$  fixed for each brane.

For non-trivial gauge group $k\geq 1$, we include the tensor product with $\gamma$ a general $k\times k$ matrix with elements $(\gamma)_{ij}=\theta_{ij}$, we then have:

\begin{displaymath}
\widehat{\theta}_{kj}=\gamma_{kj} \otimes \sum_{\alpha,\beta}E_{(\alpha,\beta) (\alpha,\beta)}q^{\alpha k}w^{\beta j}= \gamma_{kj}\otimes \widetilde{\Theta_{kj}}
\end{displaymath}

Using the  composition rule given in equation (\ref{eq:Ecomm}), we have that

\begin{align}
[E_X,\widetilde{\Theta}_{kj}]&=(q^{k}w^{sj}-1)\widetilde{\Theta}_{kj} E_ X\\
[E_Y,\widetilde{\Theta}_{kj}]&=(q^{lk}w^{rj}-1)\widetilde{\Theta}_{kj}E_Y\\
[E_Z,\widetilde{\Theta}_{kj}]&=(q^{-(1+l)k}w^{-(s+r)j}-1)\widetilde{\Theta}_{kj}E_Z
\end{align}

If we denote $\Lambda^i=diag(\lambda_1^i,\lambda_2^i,\ldots,\lambda_k^i)$ and note the fact that for $n\times m$ $a,b$  and $k\times k$ $A,B$ we have $(A\otimes a)(B \otimes b)=(AB)\otimes(ab)$ we can compute the commutators of $\widehat{X}$, $\widehat{Y}$ and $\widehat{Z}$ with $\widehat{\Theta}_{kj}$ for non-trivial gauge groups.  Using $X$ as an example we find matrices of the form

\begin{equation}
[X,\widehat{\Theta}_{kj}] =
\left( \begin{array}{ccc}
\theta_{11}\lambda_1^1[E_X,\widetilde{\Theta}_{kj}] & \theta_{12}(\lambda_1^1E_X\widetilde{\Theta}_{kj} -\lambda_2^1\widetilde{\Theta}_{kj}E_X) & \ldots \\
\theta_{21}(\lambda_2^1E_X\widetilde{\Theta}_{kj} -\lambda_1^1\widetilde{\Theta}_{kj}E_X) & \theta_{22}\lambda_2^1[E_X,\widetilde{\Theta}_{kj}] & \ldots \\
\vdots & \vdots & \ddots
\end{array} \right)
\end{equation}
 One finds that in general the variations take the form:

\begin{eqnarray}
\delta{X}&=&\theta_{ab}^{kj}(\lambda_a^1q^kw^{sj}-\lambda_b^1)\Gamma_{ab}\otimes \widetilde{\Theta}_{kj}E_X\\
\delta {Y}&=&\theta_{ab}^{kj}(\lambda_a^2q^{lk}w^{rj}-\lambda_b^2)\Gamma_{ab}\otimes \widetilde{\Theta}_{kj}E_Y\\
\delta Z&=&\theta_{ab}^{kj}(\lambda_a^3q^{-(1+l)k}w^{-(s+r)j}-\lambda_b^3)\Gamma_{ab}\otimes \widetilde{\Theta}_{kj}E_Z
\end{eqnarray}

Here the matrix $\Gamma_{ab}$ has elements $(\Gamma_{ab})_{cd}=\delta_{ac}\delta_{bd}$ and the $\theta$'s have added super-indices to note their dependence on the $kj^{th}$ generator.

In order to find the correct form of the measure we calculate the volume of these orbits, $Tr(
\delta\widehat{X}^{\dagger}\delta\widehat{X} + \delta\widehat{Y}^{\dagger}\delta\widehat{Y} +\delta\widehat{Z}^{\dagger}\delta\widehat{Z})$.  We note that $(A\otimes B)^{\dagger} = A^{\dagger}\otimes B^{\dagger}$ and that for matrices of the form $E_{(\alpha,\beta)(\gamma,\delta)}$ we expect $E^{\dagger}_{\phantom{\dagger}( \alpha,\beta)(\gamma,\delta)} = E_{(\gamma,\delta)(\alpha,\beta)}$.  After all of this,  we find that:

\begin{eqnarray}
\tr(\delta X^{\dagger} \delta X) &\sim & \sum\vert \lambda_a^1q^kw^{sj}-\lambda_b^1\vert^2 (\delta\theta_{kj}) ^{a}_{b}\delta (\theta_{-k,-j})^{b}_{a}\\
\tr(\delta Y^{\dagger} \delta Y) &\sim&  \sum\vert \lambda_a^2q^{lk}w^{rj}-\lambda_b^2 \vert^2 (\delta\theta_{kj}) ^{a}_{b}\delta (\theta_{-k,-j})^{b}_{a}\\
\tr(\delta Z^{\dagger} \delta Z) &\sim&  \sum \vert \lambda_a^3q^{-(1+l)k}w^{-(s+r)j}-\lambda_b^3 \vert^2 (\delta\theta_{kj}) ^{a}_{b}\delta (\theta_{-k,-j})^{b}_{a}
\end{eqnarray}
where we have made all indices explicit, and we sum over $a,b,k, j$

The sum necessarily has to have a form where $(\delta\theta_{kj}) ^{a}_{b}\delta (\theta_{-k,-j})^{b}_{a}$ are paired with opposite quantum numbers. This is how the invariance of the hamiltonian with respect to the quantum symmetry is realized. Also, the pairing of the $a,b$ indices is also required by the $U(1)^k$ residual gauge invariance of the configuration. The fixed indices with respect to a diagonal basis correspond to different charges under the $U(1)^k$ symmetry.

This also implies that there is no term of the form $\delta x \delta \theta$ that mixes eigenvalue and angular variations. To see this, notice that  $\delta x$ does not transform under the quantum symmetry. Thus any non-trivial such term would have to involve $\widetilde\Theta_{00}$. But one can see that restricting to these terms one has nothing other than the ${\cal N}=4$ SYM lagrangian dynamics, where this mixing does not happen.

 From this form of the metric for the angular variables, we reproduce exactly 
the expected formula (\ref{eq:images}). Our results show that after some long manipulations we can recover the expected result by the method of images. Considering that the same formula was found in some cases with discrete torsion \cite{BCorr}, we believe the result is robust
for all orbifolds with or without discrete torsion, although for some more general cases it will probably be very complicated to disentangle all the phases and diagonalize the angular variations by the pedestrian methods used above. Presumably one can arrive at this result using more sophisticated group theory manipulations for the general case.

We should note that the calculations we did were done using a gauge group of the form $\prod U(k)$. In general, for the conformal field theory, one should use a group of the form $\prod SU(k)$ instead. The calculation for the $U(k)$ theory is simpler.
Making this change will not modify the off-diagonal components associated to $a\neq b$. 
Thus most of our calculation will be unchanged.
This modification can only affect the diagonal factor in the determinant. This is, the term
of the form
\begin{equation}
\prod_i \prod_{\gamma\in \Gamma, \gamma\neq 1}|\vec x_i -\vec x^\gamma_i|^2
\end{equation}
The main result of the use of the method of images between different branes still holds. There can be a small correction to the effective repulsion between a brane and it's own images, but this is subleading in  $1/N$ and becomes unimportant in the thermodynamic limit that defines the emergent geometry. Calculating the exact answer for the $SU(k)^{nm}$ case seems rather hard in general, as one needs to diagonalize a complicated quadratic form with $nm(k-1)$ variables from the Cartan elements of each $SU(k)$. 

\section*{Acknowledgements} D. B. would like to thank D. Correa and S. Vazquez for discussions related to this work. The work of D.B. is supported in part by DOE grant DE-FG01-91ER40618. D. B. would also like to thank the Weizmann institute of science  for their hospitality while this work was being pursued.

\end{document}